\documentclass[singlespacing]{elsart}

\usepackage{graphicx}

\usepackage{amssymb}
\journal{Physica A}
\begin{document}

\begin{frontmatter}

\title{Exact solution of Smoluchowski's equation for reorientational motion in Maier-Saupe potential}

\author{A.E. Sitnitsky},
\ead{sitnitsky@mail.knc.ru}

\address{Institute of Biochemistry and Biophysics, P.O.B. 30, Kazan
420111, Russia. e-mail: sitnitsky@mail.knc.ru }

\begin{abstract}
The analytic treatment of the non-inertial rotational diffusion equation, i.e., of the Smoluchowski's one (SE), in a symmetric genuinely double-well Maier-Saupe uniaxial potential of mean torque is considered. Such potential may find applications to reorientations of the fragments of structure in polymers and proteins. We obtain the exact solution of SE via the confluent Heun's function. The solution is uniformly valid for any barrier height. We apply the obtained solution to the calculation of the mean first passage time and the longitudinal correlation time and obtain their precise dependence on the barrier height. In the intermediate to high barrier (low temperature) region the results of our approach are in full agreement with those of the approach developed by Coffey, Kalmykov, D\'ejardin and their coauthors. In the low barrier (high temperature) region our results noticeably distinguish from the predictions of the literature formula and give appreciably greater values for the transition rates from the potential well. The reason is that the above mentioned formula is obtained in the stationary limit. We conclude that for very small barrier heights the transient dynamics plays a crucial role and has to be taken into account explicitly. When this requirement is satisfied (as, e.g, at the calculation of the longitudinal correlation time) we obtain absolute identity of our results with the literature formula in the whole range of barrier heights. The drawbacks of our approach are its applicability only to the symmetric potential and its inability to yield an analytical expression for the smallest non-vanishing eigenvalue.

\end{abstract}

\end{frontmatter}

\section{Introduction}
Rotational reorientations are a particular type of motion whose utmost importance for applications can hardly be underestimated. They are ubiquitous in physics-chemical studies of liquid crystals, polymers, proteins, lipids and many other kinds of the stuff. Rotational reorientations are investigated experimentally with the help of NMR, dielectric relaxation spectroscopy, fluorescence depolarization, etc (see, e.g., \cite{Bri96} and refs. therein). The main theoretical tool for their investigation is the Smoluchowski's equation (SE), i.e., non-inertial rotational diffusion equation for a rigid body in an external potential of mean torque. It is an approximation to the more general master equation (ME). The latter is one of the basic tools in non-equilibrium statistical mechanics and physical kinetics \cite{Kam07}, \cite{Gar85}, \cite{Opp77}. ME is that for the time evolution of the probability density and consequently it expresses the fundamental principle of kinetic balance. Continuous ME is an integro-differential equation and generally it is rather difficult for analytical treatment. ME is the starting point for all models of molecular reorientation \cite{Vol94}. In each particular case physical intuition and/or first principles calculations have to be used in order to formulate an explicit expression for the transition probability which determines the entire process. However, in only very few examples the transition probability allows an analytic solution of ME. This difficulty is the reason why (instead of looking for an analytic solution) one usually tries to find approximations to the original ME. The well-known Kramers-Moyal expansion, e.g., transforms ME into a partial differential equation of infinite order. The transition from ME (by Taylor series expansion of the jump probabilities and the probability density for infinitely small jump steps) results in the Fokker-Planck equation which is a partial differential one \cite{Kam07}, \cite{Gar85}, \cite{Ris89}. For the case when the inertial effects are negligible (overdamped limit) the latter is reduced to SE. There is vast literature on their study and applications \cite{Kam07}, \cite{Gar85}, \cite{Ris89}, \cite{Cof01}, \cite{Fav60}, \cite{Fre64}, \cite{Nor70}, \cite{Nor72}, \cite{Pol73}, \cite{Val73}, \cite{Fre77}, \cite{Kun78}, \cite{Zan83}, \cite{Cof84}, \cite{Vol87}, \cite{Han90}, \cite{Tar91}, \cite{Ber93}, \cite{Pol93}, \cite{Ale00}, \cite{Fel02}, \cite{Cof04}, \cite{Cof06}, \cite{Kal09}, \cite{Kal11}. The applications of them to the dielectric spectroscopy \cite{Mar71}, \cite{Sto85}, \cite{Cof06}, fluorescence depolarization \cite{Zan83} and NMR relaxation of liquid crystals (see \cite{Don97}, \cite{Don02}, \cite{Don10} and refs. therein) have been thoroughly explored. Also, the extension of the theory from ordinary diffusion to the fractional one is intensively studied (see \cite{Cof04}, and refs. therein).

As mentioned above, SE is only an approximation to ME valid for the case of infinitely small jump steps in a space variable. According to \cite{Vol94} all models for molecular motion are divided into a jump process and a diffusion process. The jumps can take place between discrete set of accessible places or in the  continuous region of accessible places. In the latter case the diffusion limit is that of infinitely small jump steps. There is a fundamental result that "a diffusion process always be approximated by a jump process, not the reverse" \cite{Gar85}. Thus the diffusion model (\cite{Fav60}, \cite{Nor70}, \cite{Nor72}, \cite{Zan83}, \cite{Tar91}, \cite{Ber93}) can be inferred from the jump model. In this regard SE is subordinate to ME. One has to resort to SE because in the general case its treatment is easier than that of ME. SE arises in the theory of ferromagnetism (where it is called Brown's equation) \cite{Cof12} and that of molecular rotational motion in a uniaxial potential \cite{Kal09}, \cite{Cof04} (occurring for dynamics of liquid crystals, dielectric relaxation, etc.). In its turn, a particular case of SE in a uniaxial double-well potential of the mean torque can be analyzed with the help of the effective potential comprising the Maier-Saupe one as an ingredient. The latter is widely used in the theory of rotational reorientations of nematic liquid crystals \cite{Don97}, \cite{Don02}, \cite{Don10}, theory of ferromagnetism \cite{Bro63}, \cite{Bro79}, \cite{Cof12} and in the theory of dielectric relaxation of rod-like molecules \cite{Cof04}, \cite{Kal09}.

The most powerful approach to the problem was developed by Coffey, Kalmykov, D\'ejardin and their coauthors in the above mentioned papers \cite{Cof01}, \cite{Cof84}, \cite{Cof04}, \cite{Cof06}, \cite{Kal09}, \cite{Kal11}. For the sake of brevity we further call it as CKD. The approach is based on the expansion of the probability distribution function as a series of spherical harmonics. This method is well suited for the potentials
of mean torque that can be expanded in terms of spherical harmonics. It results in an infinite hierarchy of differential-recurrence relations for the the moments (the expectation values of the
spherical harmonics). CKD is an exact approach that uses no approximations and imposes no physical limitations. For instance, the exact solution in the form of the Green function in the frequency domain was given as a continued fraction \cite{Cof94}. Moreover, the correlation time, mean first passage time, etc. may be rendered exactly by writing the continued fractions in integral form. Also, the finite integral representations of the various relaxation times may be calculated directly by quadratures \cite{Cof98}, \cite{Cof961}, \cite{Cof001}. However, CKD still leaves room for complementary development that uses the expansion of the solution over eigenfunctions of Smoluchowski's operator rather than that over Legendre polynomials.

In the present paper we deal with one-dimentional motion along the polar angle $\psi$ coordinate. Such situation can arise in two ways. First, the system is invariant relative the rotations along the azimuthal angle $\varphi$ coordinate. This case is familiar in the theory of nematic liquid crystals, that of ferromagnetism, etc. and is well described by the ordinary Maier-Saupe uniaxial potential of mean torque having the minimums at $\psi=0$ and $\psi=\pi$. Many well known results exist for this case in the literature (see above cited papers).
The normalization of the probability distribution function for this case is given below (\ref{eq1}).
Second, the system undergoes reorientations along the polar angle $\psi$ coordinate at fixed azimuthal angle $\varphi=\varphi_0=const$. Such case can take place for the motion (in molecular frame) of fragments of structure in polymers or in proteins. For instance, the motion of $\Omega$-loops in proteins (see, e.g., \cite{Xia01}, Sec. III A in  \cite{Kok12} and refs. therein) may be considered as a reorientation between stable conformations one of which is "open" state and the other is "closed" one. Such motion has important functional role because the $\Omega$-loop usually plays the role of a lid that regulates the penetration of a ligand into protein interior or the access of a substrate into enzyme active site as, e.g., in the case of triosephosphate isomerase \cite{Xia01}. These reorientations between stable conformations may be considered as a transition between the wells of the corresponding potential over its barrier. It should be stressed that the minimums of the potential in this case do not necessarily lie at $\psi=0$ and $\psi=\pi$ and we need a genuinely double-well potential. A convenient way to obtain such potential is to add a logarithmic contribution to the Maier-Saupe one. Such contribution is suggested by the potential devised in the paper of Pastor and Szabo \cite{Pas92}. There may be different suitable functions under the logarithm but we show that the one used in the present paper is the mostly convenient one because it makes it possible to obtain the representation of the eigenfunctions of the Smoluchowski's operator via the known and rather well studied (up to the point that it is tabulated in Maple) special function, namely the confluent Heun's function (CHF). The resulting double-well potential for reorientational motion is a model typical one analogous to the famous $-a x^2+b x^4$ for translational motion. Unfortunately, there is a tiny problem. The normalization of the probability distribution function for this case is
\[
\frac{1}{4\pi}\int \limits_{0}^{2\pi} d\varphi\ \delta \left(\varphi-\varphi_0\right)
\ \int \limits_{0}^{\pi} d\psi
\ \sin \psi \ f(\psi, t)=\frac{1}{4\pi}\int \limits_{0}^{\pi} d\psi
\ \sin \psi \ f(\psi, t)=1
\]
In all other respects the description of the motion along the polar angle $\psi$ coordinate remains absolutely the same as in the first case. Thus, the results for two realistic cases can not be directly compared with one another because it is necessary to keep in mind some "transformation" constant. We find it awkward to suggest a model for the second case that can not be directly compared with the previous literature data for the first case. For this reason we construct a combined model potential. We use the double well Maier-Saupe potential of the second case for the polar angle $\psi$ but at the same time use the isotropy over the azimuthal angle $\varphi$ of the first one (i.e., make use of the normalization (\ref{eq1})). The resulting uniaxial potential of mean torque may seem unphysical but it has the advantage that it provides direct comparison of our results with the known literature ones especially with those of CKD approach. We show that the approach developed in the present paper is valid for such hypothetical potential and hence it remains robust for a realistic situation of the second case.

Thus, it seems interesting to obtain the analytic solution of the problem making use the expansion over eigenfunctions of Smoluchowski's operator for the double well Maier-Saupe potential that makes it possible and to compare the results of CKD with it. In the present paper we show that for the case of symmetric potential the problem under consideration appears to be amenable to stringent analytic treatment. We obtain the exact solution for the case of SE for reorientational motion in a symmetric double-well Maier-Saupe uniaxial potential of mean torque via CHF. Our solution is uniformly valid for any barrier height. The CHF is a known and by now well described special function which is a solution of the confluent Heun's equation \cite{Ron95}, \cite{Sla00}, \cite{Fiz12}, \cite{Fiz10}. We apply the obtained solution to the calculation of the mean first passage time (MFPT) and the longitudinal correlation time and obtain their precise dependence on the barrier height.

The paper is organized as follows.  In Sec. 2 the problem under study is formulated.  In Sec. 3 the solution of SE is presented. In Sec. 4 the probability distribution function is obtained. In Sec. 5 the general result is exemplified by the calculation of the escape rate from a well in the double-well Maier-Saupe uniaxial potential of mean torque. Also the longitudinal correlation time is calculated. In Sec. 6 the results are discussed and the conclusions are summarized. In Appendix the numerical calculations with CHF are considered.

\section{Smoluchowski's equation}
In the spherical frame $\Xi\equiv \{\psi, \varphi\}$ (where $ 0 \leq \psi \leq \pi$ is the polar angle and $0 \leq \varphi \leq 2\pi$ is the azimuthal one) we introduce the conditional probability $P\left(\Xi,t;\Xi_0,0\right)$ of finding the probe at orientation $\Xi$ at time $t$, if the orientation was $\Xi_0$ at time zero and the equilibrium distribution function $P_{eq}\left(\Xi\right)$. The latter represents the equilibrium probability of finding the probe at orientation $\Xi$ and is connected to the anisotropic potential of mean torque $U\left(\Xi\right)$ through the Boltzmann distribution.
In many practical situations the distribution function $P\left(\Xi,t;\Xi_0,0\right)$ and $P_{eq}\left(\Xi\right)$ characterizing a system of interest usually depend only on the polar angle $\psi$ (that between the axis of the probe and the $z$ axis of the chosen frame). The most notable example is nematic liquid crystals in a uniaxial phase for which a rotation about the director, assumed to be the $z$ laboratory axis, should leave the system invariant. Such situations arise for axially symmetric potentials when there are no dynamical coupling between the longitudinal and the transverse modes of motion. In this case the longitudinal modes are governed by the single state variable (polar angle $\psi$ that is called colatitude \cite{Cof12}) while the azimuthal angle gives rise only to a steady precession round the axis $z$. As is stressed in \cite{Cof12} in the theory of ferromagnetism the exact Fokker-Planck equation in the single variable (i.e., SE) follows directly from the axial symmetry of the potential. In the theory of non-inertial rotational diffusion for a rigid body the requirement of strong damping is necessary besides the above one \cite{Cof12}.

For the sake of brevity we further denote such distribution functions as
\[
P\left(\Xi,t;\Xi_0,0\right)=P\left(\psi,t;\psi_0,0\right)\equiv f(\psi, t)
 \]
and $P_{eq}\left(\Xi\right)=P_{eq}\left(\psi\right)\equiv f_{eq}(\psi)=f(\psi, t \rightarrow \infty)$. The latter is defined by the potential of mean torque $V(\psi)$ originating from the long range order of the system
\[
f_{eq}(\psi)=const\ \exp\left[-V(\psi)/\left(k_BT\right)\right]
\]
The distribution function $f(\psi, t)$ must be  a solution of ME or approximately of the corresponding SE. The latter is the equation for the time evolution of the distribution function $f(\psi, t)$. This function must be normalized so that its integral over the whole space gives
\begin{equation}
\label{eq1} \frac{1}{4\pi}\int \limits_{0}^{2\pi} d\varphi
\ \int \limits_{0}^{\pi} d\psi
\ \sin \psi \ f(\psi, t)=\frac{1}{2}\int \limits_{0}^{\pi} d\psi
\ \sin \psi \ f(\psi, t)=1
\end{equation}
The normalized initial condition takes the form
\begin{equation}
\label{eq2} f(\psi, 0)=\frac{2}{\sin \psi_0}\ \delta (\psi-\psi_0)
\end{equation}

For a uniaxial potential of mean torque $V(\psi)$ SE under consideration is \cite{Kal09}, \cite{Cof04}
\[
2\tau\frac{\partial f(\psi,t)}{\partial t}=\frac{\beta}{\sin \psi}
\frac{\partial}{\partial \psi}\left[\sin \psi f(\psi,t)\frac{\partial V(\psi)}{\partial \psi}\right]+
\]
\begin{equation}
\label{eq3}
\frac{1}{\sin \psi}\frac{\partial}{\partial \psi}\left[\sin \psi \frac{\partial f(\psi,t)}{\partial \psi}\right]
\end{equation}
where $\tau$ is the characteristic relaxation time for isotropic non-inertial rotational diffusion (e.g., $\tau_D$ the Debye one for the theory of dielectric relaxation or $\tau_N$ the N\'eel one for the theory of ferromagnetism) and $\beta=1/\left(k_BT\right)$. We introduce a new variable
\begin{equation}
\label{eq4} x=\cos\psi
\end{equation}
Then the equation takes the form
\begin{equation}
\label{eq5} \frac{\partial f(x,t)}{\partial t}=\frac{1}{2\tau}
\frac{\partial}{\partial x}\left[(1-x^2)\left(\frac{\partial f(x,t)}{\partial x}+\beta f(x,t)V'(x)\right)\right]
\end{equation}
where the dash means the derivative over variable $x$.

In the stationary limit $t\rightarrow \infty$ the probability distribution function $f(x, t)$ must tend to its equilibrium value $f_{eq}(x)$. Thus we have
\[
 \frac{\partial f(x, t)}{\partial t}=0
\]
and
\begin{equation}
\label{eq6}  f(x, t\rightarrow \infty)\rightarrow f_{eq}(x)=const\ \exp\left(-\frac{V(x)}{k_BT}\right)
\end{equation}

Further we consider the effective double-well potential with logarithmic contribution $-\frac{1}{\beta}\ln (1-x^2)$. This form is suggested by the potential devised in \cite{Pas92}
\begin{equation}
\label{eq7} V(x)=U(x)-\frac{1}{\beta}\ln (1-x^2)
\end{equation}
where the ingredient $U(x)$ is responsible for the barrier. The Maier-Saupe contribution is most widely used in the literature
\begin{equation}
\label{eq8} U(x)=U_{MS}(x)=b(1-x^2)
\end{equation}
where the parameter $b$ defines the barrier height. We introduce the dimensionless parameter
\begin{equation}
\label{eq9} \alpha=\frac{\beta b}{2}
\end{equation}
The obtained double-well Maier-Saupe uniaxial potential of mean torque is depicted in Fig.1.
\begin{figure}
\begin{center}
\includegraphics* [width=\textwidth] {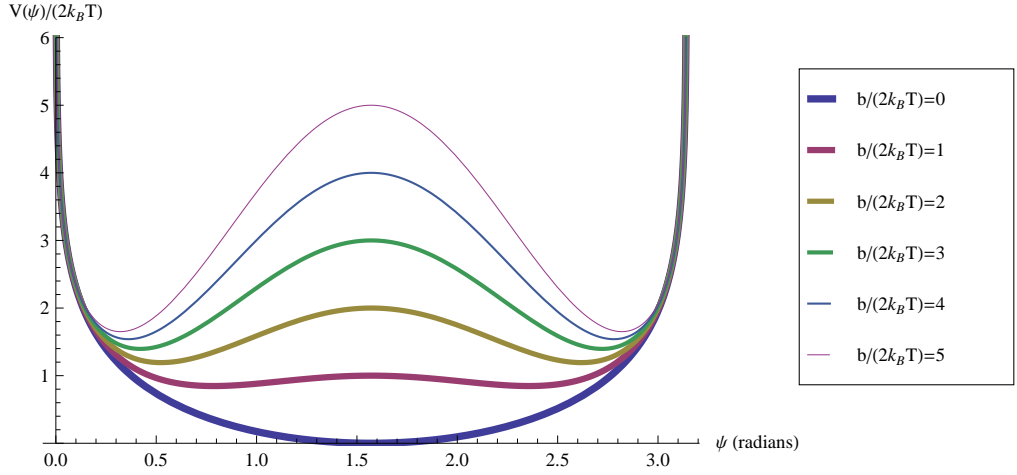}
\end{center}
\caption{The symmetric double-well Maier-Saupe uniaxial potential of mean torque $V(\psi)=b\ \sin^2 \psi-k_B T\ ln \left (\sin^2 \psi\right)$. It is called the double-well potential in the present paper to oppose it to the ordinary Maier-Saupe one $U_{MS}(\psi)=b\ \sin^2 \psi$. The barrier height is $\sigma =b/(k_B T)-1-\ln \left[b/(k_B T)\right]$.} \label{Fig.1}
\end{figure}
We call this potential the double-well Maier-Saupe to distinguish it from the ordinary Maier-Saupe one $U_{MS}(x)$.

\section{Solution of Smoluchowski's equation}
Equation (\ref{eq5}) can be solved by separation of variables
\begin{equation}
\label{eq10} f(x, t)=\eta(x)\chi(t)
\end{equation}
Denoting the separation constant as $-\rho$ ($\rho >0$) we obtain
\begin{equation}
\label{eq11} \chi(t)=\exp\left(-\rho t\right)
\end{equation}
and the equation for the function $\eta(x)$
\begin{equation}
\label{eq12} \left(1-x^2\right)\left[\eta''_{xx}(x)-4 \alpha x\eta'_{x}(x)\right]+2\left(\rho \tau+1-2\alpha+6\alpha x^2\right)\eta (x)=0
\end{equation}
We introduce a new variable
\begin{equation}
\label{eq13} y=x^2
\end{equation}
The equation takes the form
\[
y(y-1)\eta''_{yy}(y)+\left[-2\alpha y^2+\left(2\alpha+\frac{1}{2}\right)y-\frac{1}{2}\right]\eta'_{y}(y)-
\]
\begin{equation}
\label{eq14}
\left[3\alpha y+\frac{1}{2}\left(\rho \tau+1-2\alpha\right)\right]\eta(y)=0
\end{equation}
It belongs to a class of the so-called confluent Heun's equation \cite{Ron95}. Equation (\ref{eq14}) has fundamental solutions that can be expressed via the confluent Heun's function (CHF). The latter is a known special function \cite{Ron95}, \cite{Fiz12}. At present it is realized explicitly in the only symbolic computational software package Maple as $HeunC$ and its derivative $HeunCPrime$ (see \cite{Fiz12} for expert opinion on the merits and drawbacks of this computational tool in Maple). The fundamental solutions of (\ref{eq14}) are
\begin{equation}
\label{eq15} \eta^{(1)}(y)=\ HeunC\left(-2\alpha,-1/2,-1,-\frac{5\alpha}{2},\frac{\alpha-\rho \tau}{2};y\right)
\end{equation}
\begin{equation}
\label{eq16}\eta^{(2)}(y)=\ y^{1/2}HeunC\left(-2\alpha,1/2,-1,-\frac{5\alpha}{2},\frac{\alpha-\rho \tau}{2};y\right)
\end{equation}
The spectrum of the eigenvalues for the parameter $\rho$ is defined by the boundary conditions. For the latter we impose the usual requirement that the distribution function $f(x,t)$ must be finite at the boundary. In fact, due to the divergence of our double-well Maier-Saupe uniaxial potential of mean torque
at the boundary ($V(x)\rightarrow \infty$ at $x\rightarrow \pm 1$) we require that $f(x,t)$ must be zero there
\begin{equation}
\label{eq17} lim_{x\rightarrow \pm 1} f(x,t)=0
\end{equation}
The latter yields
\begin{equation}
\label{eq18} lim_{x\rightarrow \pm 1} \eta^{(i)} (x)=0
\end{equation}
for both $i=1,2$ or equivalently
\begin{equation}
\label{eq19} lim_{y\rightarrow 1} \eta^{(i)} (y)=0
\end{equation}
We denote
\begin{equation}
\label{eq20} \Delta_n=\rho_n^{(1)} \tau
\end{equation}
\begin{equation}
\label{eq21} \Omega_m=\rho_m^{(2)} \tau
\end{equation}
The equation for the spectrum of eigenvalues of $\Delta_n$ ($n=1,2,3,...$) for $\eta^{(1)}_n(y)$ is
\begin{equation}
\label{eq22}   HeunC\left(-2\alpha,-1/2,-1,-\frac{5\alpha}{2},\frac{\alpha-\Delta_n}{2};y\rightarrow 1\right)=0
\end{equation}
That for the spectrum of eigenvalues of $\Omega_m$ ($m=1,2,3,...$) for $\eta^{(2)}_m(y)$ is
\begin{equation}
\label{eq23}   HeunC\left(-2\alpha,1/2,-1,-\frac{5\alpha}{2},\frac{\alpha-\Omega_m}{2};y\rightarrow 1\right)=0
\end{equation}
These equations can be solved only numerically but Maple easily copes with this problem.
One always obtains $\Delta_1=0$ while the lowest $\Omega_1$ is always nonzero and the corresponding term mainly determines the behavior of the distribution function $f(\psi,t)$ and MFPT calculated with its help. The dependence of $\Omega_n$ and $\Delta_m=0$ on the parameter $\alpha$ is depicted in Fig.2 and Fig.3 respectively.
\begin{figure}
\begin{center}
\includegraphics* [width=\textwidth] {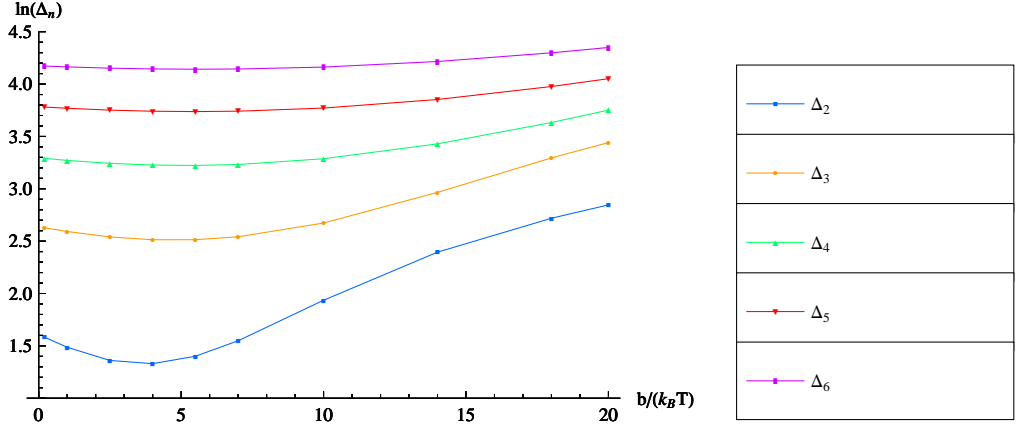}
\end{center}
\caption{The dependence of the spectrum of the eigenvalues of $\Delta_n$ (obtained as the solution of  (\ref{eq22})) on the barrier height ($\Delta_2$ is the bottom line, ..., $\Delta_6$ is the top one). This spectrum characterizes the first set of eigenfunctions (\ref{eq15}) of the Smoluchowski's operator $L_S$ if the Smoluchowski's equation under study (\ref{eq3}) to be written in the formal way $\dot f = L_S f$.}
\label{Fig.2}
\end{figure}
\begin{figure}
\begin{center}
\includegraphics* [width=\textwidth] {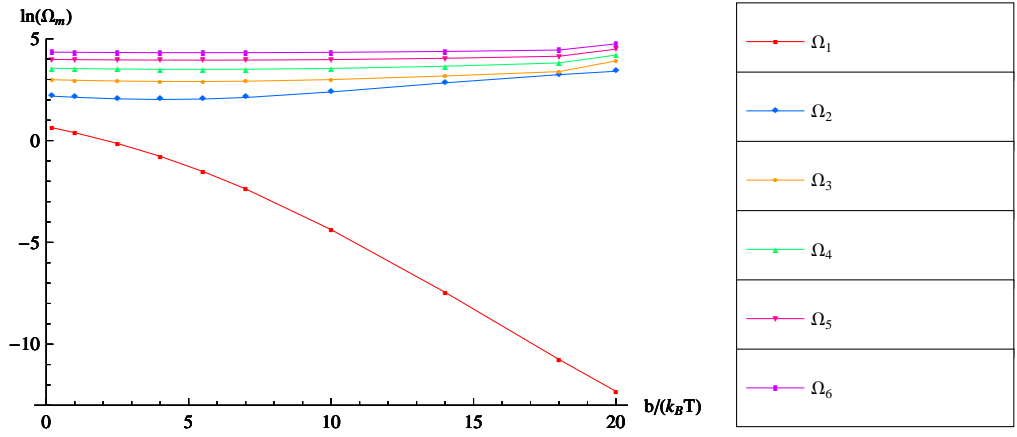}
\end{center}
\caption{The dependence of the spectrum of the eigenvalues of $\Omega_m$ (obtained as the solution of  (\ref{eq23})) on the barrier height ($\Omega_1$ is the bottom line, ..., $\Omega_6$ is the top one). This spectrum characterizes the second set of eigenfunctions (\ref{eq16}) of the Smoluchowski's operator $L_S$ if the Smoluchowski's equation under study (\ref{eq3}) to be written in the formal way $\dot f = L_S f$.}
\label{Fig.3}
\end{figure}
It should be stressed that at large barrier heights the value of $\Omega_1$ becomes very small as Fig.3 testifies.

\section{Probability distribution function}
We return to the variable $\psi$ with the help of (\ref{eq4}) and (\ref{eq13}). Then the general solution of SE (\ref{eq3}) can be written as
\[
f(\psi,t)=C_1\eta^{(1)}_1(\psi)+\sum_{n=2}^{\infty}C_n \exp \left(-\Delta_n t/\tau\right)\eta^{(1)}_n(\psi)+
\]
\begin{equation}
\label{eq24}\sum_{m=1}^{\infty}D_m \exp \left(-\Omega_m t/\tau\right)\eta^{(2)}_m(\psi)
\end{equation}
The crucial issue for the calculation of the coefficients $C_n$ and $D_m$ is the following one: being the solution of the boundary problems both $\eta^{(1)}_n(\psi)$ and $\eta^{(2)}_m(\psi)$ are full sets of the orthogonal functions. Besides, as they belong to different fundamental solutions they are orthogonal to one another
\begin{equation}
\label{eq25} \int \limits_{0}^{\pi} d\psi
\ \sin \psi\ \eta^{(i)}_l(\psi)\eta^{(j)}_m(\psi)\propto\delta_{lm}\delta_{ij}
\end{equation}
for $i,j=1,2$ and any function can be expanded into a series over $\eta^{(1)}_n(\psi)$ or $\eta^{(2)}_m(\psi)$.
At time $t=0$  (\ref{eq24}) yields (with taking into account (\ref{eq2}))
\begin{equation}
\label{eq26}\frac{2}{\sin \psi_0}\ \delta (\psi-\psi_0)=\sum_{n=1}^{\infty}C_n \eta^{(1)}_n(\psi)+
\sum_{m=1}^{\infty}D_m \eta^{(2)}_m(\psi)
\end{equation}
Then the coefficients $C_n$ and $D_m$ are obtained by multiplying (\ref{eq26}) by $\eta^{(1)}_n(\psi)$ or $\eta^{(2)}_m(\psi)$ and integration over $\psi$ with taking into account (\ref{eq25})
\begin{equation}
\label{eq27} C_n=2\eta^{(1)}_n(\psi_0)\left[\int \limits_{0}^{\pi} d\psi
\ \sin \psi\ \left[\eta^{(1)}_n(\psi)\right]^2\right]^{-1}
\end{equation}
and
\begin{equation}
\label{eq28} D_m=2\eta^{(2)}_m(\psi_0)\left[\int \limits_{0}^{\pi} d\psi
\ \sin \psi\ \left[\eta^{(2)}_m(\psi)\right]^2\right]^{-1}
\end{equation}

Thus, we have the explicit algorithm for obtaining the coefficients $C_n$ and $D_m$ along with the corresponding spectra $\Delta_n$ and $\Omega_m$ that makes the solution of our problem to be completed. Substitution of all these values into (\ref{eq24}) yields the required probability distribution function.
Of course, in practice a truncation of the series is necessary, i.e., replacement of the infinity by some finite number $M$ that is determined by the required accuracy. Below we prove that in practice the first terms $M=1$ (i.e., those with the coefficients $C_1$ and $D_1$) provide sufficient accuracy (see Fig. 6 and discussion in Sec. 6).

\section{Mean first passage time and longitudinal correlation time}
To exhibit how the result obtained may be useful we calculate with its help the dependence of MFPT on the barrier height $\alpha$. Our effective Maier-Saupe uniaxial potential of mean torque is (see (\ref{eq7}), (\ref{eq8}) and (\ref{eq9}))
\begin{equation}
\label{eq29} \frac{1}{k_B T}V(\psi)=2\alpha\ \sin^2 \psi-ln \left (\sin^2 \psi\right)
\end{equation}
It is depicted in Fig. 1. It has two local minima the left of which is at
\begin{equation}
\label{eq30} \psi_L=\arcsin \left(\frac{1}{\sqrt {2 \alpha}}\right)
\end{equation}
and the maximum at
\begin{equation}
\label{eq31}  \psi_M=\frac{\pi}{2}
\end{equation}
First we calculate the non-averaged MFTP.
We set the initial condition to be the system in the left local minimum $\psi_0=\psi_L$. It is worthy to recall that our solution of SE $f(\psi,t)$ is actually the conditional probability $f(\psi,t) \equiv P\left(\psi,t;\psi_0,0\right)$ of finding the system at orientation $\psi$ at time $t$, if the orientation was $\psi_0$ at time zero. Then, following Risken \cite{Ris89} we introduce the probability $\Omega(\psi_0, t)$ of realizations which have started at $\psi_0=\psi_L$ and which have not yet reached one of the boundaries $\psi=0$ or $\psi=\psi_M$ up to the time $t$
\begin{equation}
\label{eq32} \Omega(\psi_0, t)=\int \limits_{0}^{\psi_M} d\psi
\ \sin \psi\ f(\psi,t)
\end{equation}
The distribution function $w(\psi_0, T)$ for the first passage time $T$ (we use the notations from \cite{Ris89} and note that this value is not to be confused with temperature which will further enter only implicitly via $\alpha$ from (\ref{eq9}) and via $1/\left (k_B T\right)=2\alpha/b$) is
\begin{equation}
\label{eq33} w(\psi_0, T)=-\frac{\partial \Omega(\psi_0, T)}{\partial T}
\end{equation}
The moments of the first passage time distribution are
\begin{equation}
\label{eq34} T_n (\psi_0)=\int \limits_{0}^{\infty} dT
\ T^n w(\psi_0, T)
\end{equation}
The value of interest for us here is the MFPT $T_1$
\begin{equation}
\label{eq35} T_1 (\psi_0)=\int \limits_{0}^{\infty} dT
\ T w(\psi_0, T)
\end{equation}
that characterizes the ability for the system to reach the barrier top $\psi_M$ starting from the left minimum $\psi_0=\psi_L$. Thus, we have to calculate the quantity
\begin{equation}
\label{eq36} T_1 (\psi_0)=-\int \limits_{0}^{\psi_M} d\psi\ \sin \psi
\int \limits_{0}^{\infty} dT\ T\ \frac{\partial f(\psi, T)}{\partial T}
\end{equation}
After straightforward calculations we obtain
\[
\frac{1}{\tau}T_1 (\psi_0)=\frac{D_1}{\Omega_1}\int \limits_{0}^{\psi_M} d\psi\ \sin \psi\ \eta^{(2)}_1(\psi)+
\]
\begin{equation}
\label{eq37}
\sum_{n=1}^{\infty}\int \limits_{0}^{\psi_M} d\psi\ \sin \psi\ \left[\frac{C_n}{\Delta_n} \eta^{(1)}_n(\psi)+\frac{D_n}{\Omega_n} \eta^{(2)}_n(\psi)\right]
\end{equation}
Finally, we calculate the averaged MFPT $<T_1>$ for the case when the initial condition is thermodynamically averaged over the left well
\[
<T_1>=\left [\int \limits_{0}^{\pi} d\psi_0\ \sin \psi_0 \exp\left(-\frac{V( \psi_0)}{k_BT}\right)\right]^{-1}
\times
\]
\begin{equation}
\label{eq38}\int \limits_{0}^{\psi_M} d\psi_0\ \sin \psi_0 \exp\left(-\frac{V( \psi_0)}{k_BT}\right)T_1 (\psi_0)
\end{equation}
It is worthy to recall here that the coefficients $C_n$ and $D_n$ are the functions of the initial condition $\psi_0$, i.e., explicitly $C_n\equiv C_n \left(\psi_0\right)$ and $D_n\equiv D_n \left(\psi_0\right)$. The results of numerical calculation of $<T_1>$ are presented in Fig.4.
\begin{figure}
\begin{center}
\includegraphics* [width=\textwidth] {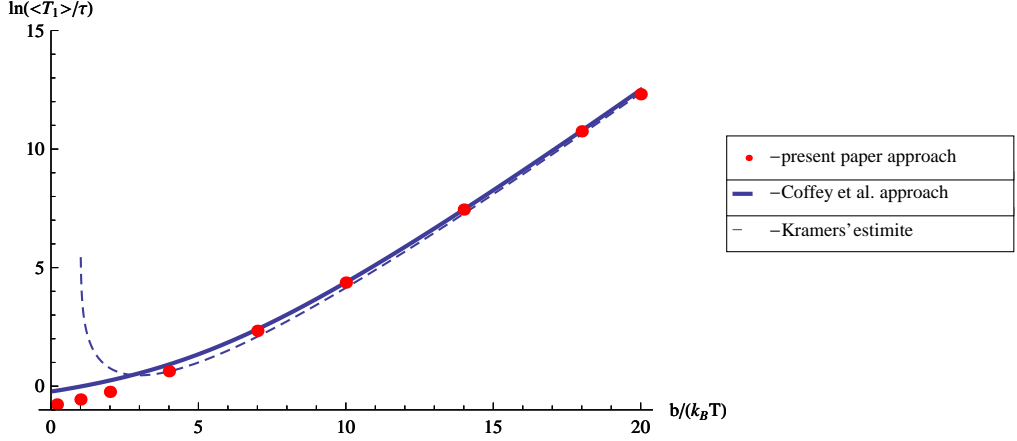}
\end{center}
\caption{The dependence of the mean first passage time for the transition from the well of the symmetric double-well Maier-Saupe uniaxial potential of mean torque $V(\psi)=b\ \sin^2 \psi-k_B T\ ln \left (\sin^2 \psi\right)$ on the parameter $b$ which characterizes the barrier height. The dots are the result of our approach developed in the present paper. The dashed line is the result of the Kramers' estimate for the transition rate in the high friction limit (\ref{eq45}). The continuous line is the result of the approach developed by Coffey, Kalmykov, D\'ejardin and their coauthors (\ref{eq42}).}
\label{Fig.4}
\end{figure}
There the results for the averaged MFPT $<T_1>$ obtained from our numerical calculations are compared with the analytical estimate from the Kramers' theory for the transition rate
\begin{equation}
\label{eq39} \frac{1}{\tau}<T_1>\approx\frac{1}{\Gamma_K}
\end{equation}
If we return again to the variable $x=\cos\psi$ and denote
\begin{equation}
\label{eq40} h(x)\equiv\frac{1}{k_B T}V(x)=2\alpha\ (1-x^2) -ln (1-x^2)
\end{equation}
then the Kramers' formula takes the form \cite{Kra40}, \cite{Han90},
\begin{equation}
\label{eq41} \Gamma_K=\frac{\sqrt{h''_{xx}\left(x_L\right)\mid h''_{xx}\left(x_M\right)\mid}}{2\pi}
\exp \left\{-\left[h\left(x_M\right)-h\left(x_L\right)\right]\right \}
\end{equation}
Also in Fig.4 the results of the approach of Coffey, Kalmykov, D\'ejardin and their coauthors (CKD) are presented. The latter in our case of the potential (\ref{eq29}) is eq. (1.18.1.7) of \cite{Cof04}
\begin{equation}
\label{eq42} \frac{<T_1>}{\tau}=2 \int \limits_{0}^{\pi/2} d\psi'\ \frac{e^{2\alpha\ \sin^2 \psi'-ln \left (\sin^2 \psi'\right)}}{\sin \psi'}\int \limits_{0}^{\psi'} d\psi\ \sin \psi\ e^{-2\alpha\ \sin^2 \psi+ln \left (\sin^2 \psi\right)}
\end{equation}

Following CKD we introduce the longitudinal correlation function
\[
\frac{<\cos \psi(0)\cos \psi(t)>}{<\cos^2 \psi(0)>}= \left\{\int \limits_{0}^{\pi} d\psi
\ \sin \psi \int \limits_{0}^{\pi} d\psi_0
\ \sin \psi_0 \ \cos \psi_0 \cos \psi\ f(\psi, t)\right\}\times
\]
\begin{equation}
\label{eq43} \left\{\int \limits_{0}^{\pi} d\psi
\ \sin \psi \int \limits_{0}^{\pi} d\psi_0
\ \sin \psi_0 \ \left(\cos \psi_0\right)^2 \ f(\psi, t)\right\}^{-1}
\end{equation}
where the averaging $<...>$ is carried out with the help of the obtained distribution function $f(\psi, t)$ given by (\ref{eq24}). Also, we define the longitudinal correlation time
\begin{equation}
\label{eq44} \tau_{\mid\mid}=\int \limits_{0}^{\infty} dt\ \frac{<\cos \psi(0)\cos \psi(t)>}{<\cos^2 \psi(0)>}
\end{equation}
In Fig.5 the results of calculations of the value $\tau_{\mid\mid}$ are presented.
\begin{figure}
\begin{center}
\includegraphics* [width=\textwidth] {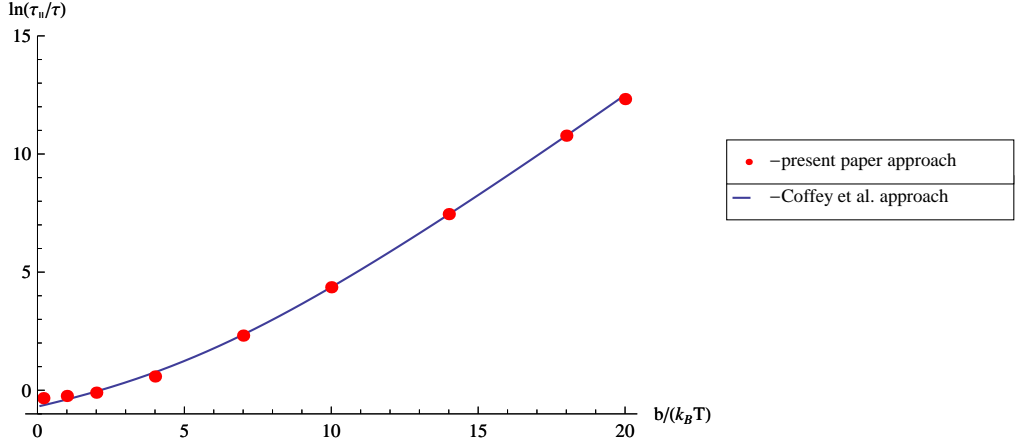}
\end{center}
\caption{The dependence of the longitudinal correlation time for rotational motion in the symmetric double-well Maier-Saupe uniaxial potential of mean torque $V(\psi)=b\ \sin^2 \psi-k_B T\ ln \left (\sin^2 \psi\right)$ on the parameter $b$ which characterizes the barrier height. The dots are the result of our approach developed in the present paper. The line is the result of the approach developed by Coffey, Kalmykov, D\'ejardin and their coauthors (\ref{eq45}).}
\label{Fig.5}
\end{figure}
They are compared with the result of CKD that in the case of the potential (\ref{eq29}) is eq. (2.10.25) of \cite{Cof04}
\[
\frac{1}{\tau}\tau_{\mid\mid}^{CKD}=\frac{2}{Z<\cos^2 \psi>_0}\times
\]
\begin{equation}
\label{eq45}  \int \limits_{0}^{\pi} d\psi'\ \frac{e^{2\alpha\ \sin^2 \psi'-ln \left (\sin^2 \psi'\right)}}{\sin \psi'}\left(\int \limits_{0}^{\psi'} d\psi\ \sin \psi\ \cos \psi\e^{-2\alpha\ \sin^2 \psi+ln \left (\sin^2 \psi\right)}\right)^2
\end{equation}
where
\begin{equation}
\label{eq46} Z=\int \limits_{0}^{\pi} d\psi\ \sin \psi\ e^{-2\alpha\ \sin^2 \psi+ln \left (\sin^2 \psi\right)}
\end{equation}
and
\begin{equation}
\label{eq47} <\cos^2 \psi>_0=\frac{1}{Z}\int \limits_{0}^{\pi} d\psi\ \sin \psi\ \cos^2 \psi\ e^{-2\alpha\ \sin^2 \psi+ln \left (\sin^2 \psi\right)}
\end{equation}
Finally, in Fig.6 we plot the dependence of our calculated MFPT on the truncation number $M$ in the series (\ref{eq24}).
\begin{figure}
\begin{center}
\includegraphics* [width=\textwidth] {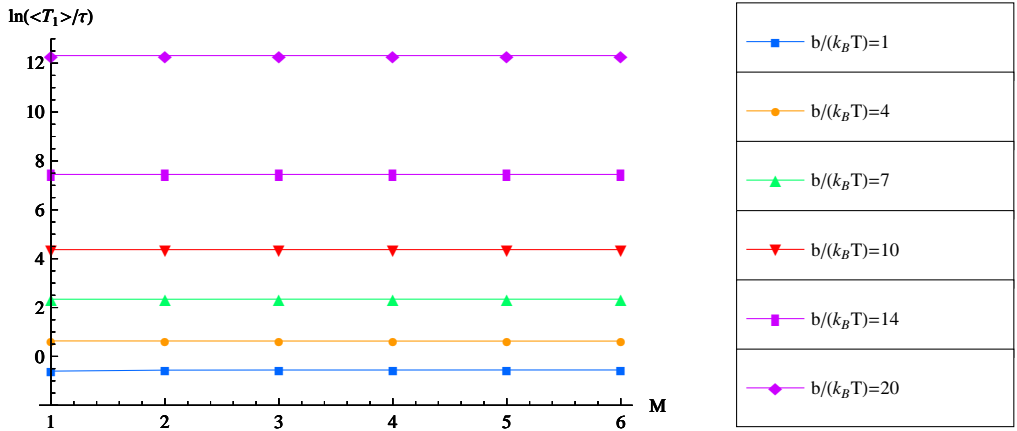}
\end{center}
\caption{The dependence of the mean first passage time for the transition from the well of the symmetric double-well Maier-Saupe uniaxial potential of mean torque $V(\psi)=b\ \sin^2 \psi-k_B T\ ln \left (\sin^2 \psi\right)$ on the truncation number $M$ in the series (\ref{eq24}) ($b/\left(k_B T\right)=1$ is the bottom line, ..., $b/\left(k_B T\right)=20$ is the top one).}
\label{Fig.6}
\end{figure}

\section{Results and discussion}
To clarify the terminology it should be stressed that
we deal with barrier heights in the energetic units (i.e., conceive it to be the ratio of its value in the natural units to the product of the Boltzmann constant and temperature). Thus, e.g., the low barrier regime may well be realized for high barriers in the natural units at sufficiently high temperatures. Namely in this sense we further use the notions of the intermediate to high barrier (low temperature) region and the low barrier (high temperature) one.

Fig.4 testifies our approach yields absolutely identical results with CKD one (\ref{eq42}) in the intermediate to high barrier region, i.e., at barrier heights greater than $b/\left(k_B T\right)\approx 4$ ($\sigma \approx 3$). The Kramers' estimate for the transition rate in the high friction limit (\ref{eq41}) yields excellent approximation for the double-well Maier-Saupe potential at $b/\left(k_B T\right)$ greater than $\approx 4$. We conclude that in the intermediate to high barrier region our results are in full agreement with the known literature ones.

In the low barrier region our results yield noticeably smaller values for MFPT than those predicted by both the Kramers' estimate (\ref{eq41}) and that of CKD (\ref{eq42}). We conclude that in this region the formula (\ref{eq42}) of CKD (or more accurately, eq. (1.18.1.7) of \cite{Cof04}) appreciably overestimates MFPT or in other words predicts too small values for the transition rates from the potential well. The reason of the discrepancy between our results for very small barrier heights for MFPT and those of CKD seen in Fig.4 is in the fact that their formula (\ref{eq42}) (or more accurately, eq. (1.18.1.7) of \cite{Cof04}) is obtained in the stationary limit. It seems that for very small barrier heights the transient dynamics plays a crucial role and has to be taken into account explicitly. The latter requirement is satisfied both in our approach and at deriving the CKD formula for the longitudinal correlation time (\ref{eq45}) (or more accurately, eq. (2.10.25) of \cite{Cof04}). Fig. 5 testifies that for the longitudinal correlation time we obtain absolute identity of our results with those of CKD (\ref{eq45}) in the whole range of barrier heights.

The characteristic feature is that both MFPT (Fig.4) and the longitudinal correlation time (Fig. 5) are smaller than the relaxation time $\tau$ for very small barrier heights. It is noticeable that the formulas of CKD also lead to such a conclusion (see Fig.4 and Fig.5). The physical sense of it may be as follows. The relaxation time $\tau$ is an average characteristic rather than a minimal possible time of the processes in the system. Its value sets the time scale for the motion proceeding in the diffusional regime. The characteristic length figuring in the relaxation time is the radius of a rotating sphere that models the physical object of interest. Thus, in itself $\tau$ is a rather rough parameter that does not take into account either the characteristic distance of the motion or the form of the potential. When the barrier height in our double-well  Maier-Saupe potential becomes very small the distance between its minimums also becomes very small (they coalesce at $\psi=\pi/2$) in contrast to the ordinary Maier-Saupe one where they still remain at $\psi=0$ and $\psi=\pi$. In this case the distance for the motion becomes commensurable or even less than the characteristic length figuring in the relaxation time. In our opinion, it is not surprising that to move a vanishingly small distance over a vanishingly small barrier the system requires lesser time than $\tau$ (MFPT$\ < \tau$).

To prove the convergence of the obtained series (\ref{eq24}) we plot in Fig.6 the dependence of the calculated value for MFPT on the truncation number $M$. This Fig. shows that $M=1$ provides sufficient accuracy because the results cease to change at further increase of $M$ for all barrier heights. This fact can be anticipated for large barrier heights where the term $M=1$ or more precisely the term with the coefficient $D_1$ overwhelmingly dominates the dynamics of the system due to very low values of the smallest non-vanishing eigenvalue $\Omega_1$ (see Fig. 3). However, our results testify that also for small barrier heights the the terms with the coefficient $C_1$ and $D_1$ in the series (\ref{eq24}) are sufficient.

Our approach yields directly the probability distribution function rather than the statistical moments (i.e., values averaged over it). We recall that CKD enables one to obtain the reformulation of the problem as the differential-recurrence relations for the longitudinal and transverse correlation functions. A systematic way for their analysis has been suggested that yields "exact analytic solutions for the longitudinal and transverse complex susceptibilities and correlation times for many problems of practical interest" \cite{Cof04}. However, in some applications (e.g., to relaxation processes in NMR) it is desirable to work namely with the distribution function taken alone. The virtue of our approach is in providing such option.

Our approach yields the exact solution of SE (\ref{eq3}) along with that of CKD. Let us formally write this equation as $\dot f = L_S f$ where $L_S$ is the Smoluchowski's operator. Then the difference of our exact solution from that of CKD can be explained as follows. The authors of CKD expand $f$ over Legendre polynomials while we expand it over the eigenfunctions of $L_S$. The latter are expressed over CHF which at present can be considered as a well tabulated special function. Efficient numerical algorithms for it are available in the literature let alone an explicit realization in Maple that makes its usage a routine problem. In fact, to deal with CHF is no harder than with Legendre polynomials. Moreover, CKD approach leads to recurrence system for the coefficients in the expansion when those with lower indexes are related to higher ones. To cope with such system is a difficult problem in itself though it is successfully solved in the above mentioned papers. In our exact solution the coefficients with different indexes in the expansion are independent of each other. However, besides the advantages of our solution over the approximate one there are drawbacks. Within the framework of CKD approach the authors of the above mentioned papers manage to obtain an analytical expression for the smallest non-vanishing eigenvalue. In our approach that is obtained as a solution of (\ref{eq23}). The latter is to be solved numerically. In the present paper we do not try to obtain an analytical expression for $\Omega_1$ from (\ref{eq23}). Also, our approach is developed only for symmetric uniaxial potential of mean torque. The crucial substitution (\ref{eq13}) can work only for potentials with even powers of $\cos \psi$ or $\sin \psi$ and can not be generalized to odd powers leading to the asymmetry of the potential.

In conclusion, we revisit a rather old problem of rotational motion in a double-well potential of mean torque. The corresponding physical process is within the framework of the escape rate theory that is more than seventy years old (if to count from the cornerstone Kramers' paper). Nevertheless, the some useful tools for its exact analytic treatment (the theory of CHF and its convenient numerical realizations) were developed relatively recently. Their application provides a new way to rederive old and well known results and yields a complementary development to them.

Acknowledgements. The author is grateful to Dr. Yu.F. Zuev
for helpful discussions. The work was supported by the grant from RFBR.

\section{Appendix}
Maple is very efficient for numerical search of the roots of (\ref{eq22}) and (\ref{eq23}). Unfortunately, the calculations of the integrals containing $HeunC$ by Maple proves to be too time consuming that makes the analysis of $<T_1>$ with the help of this program to be practically inconvenient. Besides, one can not control the processing at its usage \cite{Fiz12}. For this reason, we prefer to employ the explicit numerical realization of the CHF. We find that the required accuracy can well be retained at reducing the number of terms in the definition of the CHF by a series. As a consequence the processing time is diminished substantially. For this purpose, we make use of the explicit representation of the CHF by a series \cite{Ron95}. It includes the three-term recurrence relation \cite{Ron95}, \cite{Fiz10} that in our case takes the form for the first solution of (\ref{eq14})
\[
\eta^{(1)}(y)=HeunC\left(-2\alpha,-1/2,-1,-\frac{5\alpha}{2},\frac{\alpha-\rho\tau}{2};y\right)=
\]
\begin{equation}
\label{eq48}\sum_{k=0}^{\infty}v^{(1)}_k\left(-2\alpha,-1/2,-1,-\frac{5\alpha}{2},\frac{\alpha-\rho \tau}{2}\right)y^k
\end{equation}
where
\begin{equation}
\label{eq49} A^{(1)}_k v^{(1)}_k=B^{(1)}_k v^{(1)}_{k-1}+C^{(1)}_k v^{(1)}_{k-2}
\end{equation}
with the initial conditions $v^{(1)}_{-1}=0$, $v^{(1)}_0=1$. Here
\begin{equation}
\label{eq50} A^{(1)}_k=1-\frac{1}{2k}
\end{equation}
\begin{equation}
\label{eq51} B^{(1)}_k=1+\frac{4\alpha-5}{2k}+\frac{4-4\alpha-2\rho \tau}{4k^2}
\end{equation}
\begin{equation}
\label{eq52} C^{(1)}_k=-\frac{2\alpha}{k^2}\left(k-\frac{1}{2}\right)
\end{equation}

For the second solution of (\ref{eq14}) we have
\[
y^{-1/2}\eta^{(2)}(y)=HeunC\left(-2\alpha,1/2,-1,-\frac{5\alpha}{2},\frac{\alpha-\rho \tau}{2};y\right)=
\]
\begin{equation}
\label{eq53}\sum_{k=0}^{\infty}v^{(2)}_k\left(-2\alpha,1/2,-1,-\frac{5\alpha}{2},\frac{\alpha-\rho \tau}{2}\right)y^k
\end{equation}
where
\begin{equation}
\label{eq54} A^{(2)}_k v^{(2)}_k=B^{(2)}_k v^{(2)}_{k-1}+C^{(2)}_k v^{(2)}_{k-2}
\end{equation}
with the initial conditions $v^{(2)}_{-1}=0$, $v^{(2)}_0=1$. Here
\begin{equation}
\label{eq55} A^{(2)}_k=1+\frac{1}{2k}
\end{equation}
\begin{equation}
\label{eq56} B^{(2)}_k=1+\frac{4\alpha-3}{2k}-\frac{\rho \tau}{2k^2}
\end{equation}
\begin{equation}
\label{eq57} C^{(2)}_k=-\frac{2\alpha}{k}
\end{equation}
To provide fast convergence of the series in (\ref{eq48}) and (\ref{eq53}) we truncate them, i.e., replace the infinity by some large but finite $N$ which we choose from the requirement that the obtained values of $<T_1>/\tau$ cease to change within the limits of the required accuracy. Under these approximations we obtain an efficient numerical algorithm for calculating the integrals with the our CHF.


\newpage
\begin{thebibliography}{00}
\bibitem{Bri96}
N.V.Brilliantov, O.P. Revokatov, Molecular motion in disordered media, Moscow University Press, Moscow, 1996.
\bibitem{Kam07}
N.G. Van Kampen, Stochastic processes in physics and chemistry, 3-d ed., Elsevier, 2007.
\bibitem{Gar85}
C. W. Gardiner, Handbook of stochastic methods for physics, chemistry and the natural sciences, 2-nd ed., Springer, 1985.
\bibitem{Opp77}
I. Oppenheim, K.E. Shuler, G.H. Weiss, Stochastic processes in chemical physics. The master equation. MIT Press, Cambrige, MA, 1977.
\bibitem{Vol94}
R.R. Vold, Deuterium NMR studies of molecular motion, in: Nuclear magnetic resonance probes of molecular motion, ed. R. Tycko, Kluwer, 1994, 27 - 86.
\bibitem{Ris89}
H. Risken, The Fokker-Plank equation, 2-nd ed., Springer, 1989.
\bibitem{Cof01}
W.T. Coffey, D.A. Garanin, D.J. McCarthy, Crossover formulas in the Kramers theory of thermally activated escape rates - application to spin systems, Adv. Chem. Phys. 117, Eds. I. Prigogine, S.A. Rice, John Wiley, 2001.
\bibitem{Fav60}
L.D. Favro, Theory of the rotational Brownian motion of a free rigid body, Phys.Rev. 119 (1960) 53-62.
\bibitem{Fre64}
J.H. Freed, Anisotropic rotational diffusion and electron spin resonance linewidths, J.Chem.Phys. 41 (1964) 2077-2083.
\bibitem{Nor70}
P.L. Nordio, P. Busolin, Electron spin resonance line shapes in partially oriented systems, J.Chem.Phys. 55 (1970) 5485-5490.
\bibitem{Nor72}
P.L. Nordio, G. Rigatti, U. Segre, Spin relaxation in nematic solvents,
J.Chem.Phys. 56 (1972) 2117-2123.
\bibitem{Pol73}
C.F. Polnaszek, G.V. Bruno, J.H. Freed, ESR line shapes in the slow-motional region: Anisotropic liquids, J.Chem.Phys. 58 (1973) 3185-3199.
\bibitem{Val73}
K.A. Valiev, E.N. Ivanov, Rotational Brownian motion, Uspechi Phys. Nauk (in Russian), 109 (1973) 31-64.
\bibitem{Fre77}
J.H. Freed, Stochastic-molecular theory of spin-relaxation for liquid
crystals, J.Chem.Phys. 66 (1977) 4183-4199.
\bibitem{Kun78}
F.M. Kuni, A.A. Melikhov, B.A. Storonkin,
Three dimensional rotational relaxation in external fields, Teor. Mat. Fiz., 34 (1978) 374-386.
\bibitem{Zan83}
C. Zannoni, A. Arcioni, P. Cavatorta, Fluorescence depolarization in liquid crystals and membrane bilayers, Chem.Phys.Lipids 32 (1983) 179-250.
\bibitem{Cof84}
W.T. Coffey, M. Evens, P. Grigolini, Molecular diffusion and spectra, Wiley, 1984.
\bibitem{Vol87}
R.R. Vold, R.L. Vold, Nuclear spin relaxation and molecular dynamics in ordered systems: Models
for molecular reorientation in thermotropic liquid crystals, J.Chem.Phys. 88 (1987) 1443-1457.
\bibitem{Kra40}
H.A. Kramers, Brownian motion in a field of force and the diffusion model of chemical reactions, Physica 7 (1940) 284-304.
\bibitem{Han90}
P. H\"anggi, P. Talkner, M. Borkovec, Fifty years after Kramers'
equation: reaction rate theory, Rev.Mod.Phys. 62 (1990) 251-341.
\bibitem{Tar91}
R. Tarroni, C. Zannoni, On the rotational diffusion of asymmetric molecules in liquid crystals, J.Chem.Phys. 95 (1991) 4550-4564.
\bibitem{Ber93}
E. Berggren, F.L. Tarroni, and C. Zannoni, Rotational diffusion of uniaxial probes in biaxial liquid crystal phases, J.Chem.Phys. 99 (1993) 6180-6200.
\bibitem{Pol93}
A. Polimeno, J.H. Freed, A many-body stochastic approach to rotational motion in liquids. In: Advances in chemical physics, ed. I. Prigogine and S.A. Rice, v. LXXXIII, Wiley, 1993.
\bibitem{Ale00}
W. Alexiewicz, Ensemble averages for Smoluchowski-Debye rotational diffusion
in the presence of a two-angle-dependent reorienting force, Chem.Phys.Lett. 320 (2000) 582-586.
\bibitem{Fel02}
B.U. Felderhof, Nonlinear response of a dipolar system with rotational diffusion to a rotating field, Phys.Rev. E66 (2002) 051503.
\bibitem{Cof04}
W.T. Coffey, Yu.P. Kalmykov, J.T. Waldron, The Langevin equation: with applications to stochastic problems in physics, chemistry and electrical engineering, 2-nd ed., World Scientific, 2004.
\bibitem{Cof06}
W.T. Coffey, Y.P. Kalmykov, B. Ouari, S.V. Titov, Rotational diffusion and orientation relaxation of rodlike molecules in a biaxial liquid crystal phase, Physica A 368 (2006) 362-376.
\bibitem{Kal09}
Y.P. Kalmykov, S.V. Titov, W.T. Coffey, Inertial effects in the orientational relaxation of rodlike molecules in a uniaxial potential, J.Chem.Phys. 130 (2009) 064110.
\bibitem{Kal11}
Y.P. Kalmykov, S.V. Titov, W.T. Coffey, Inertial and bias effects in the rotational Brownian motion of rodlike molecules in a uniaxial potential, J.Chem.Phys. 134 (2011) 044530.
\bibitem{Mar71}
A.J. Martin, G. Meier, A. Saupe, Extended Debye theory for dielectric relaxation in nematic liquid crystals, Symp. Faraday Soc., 5 (1971) 119-133.
\bibitem{Sto85}
B.A. Storonkin, Theory of dielectric relaxation in nematic liquid crystals, Kristallogr., 30 (1985) 841-853.
\bibitem{Don97}
R.Y. Dong, Nuclear magnetic resonance of liquid crystals, Springer Verlag,
1997.
\bibitem{Don02}
R.Y. Dong, Relaxation and the dynamics of molecules in the liquid
crystalline phases, Progress in Nuclear Magnetic Resonance Spectroscopy 41 (2002) 115-151.
\bibitem{Don10}
R.Y. Dong, Nuclear magnetic resonance spectroscopy of liquid crystals, World
Scientific Publishing Co., 2010.
\bibitem{Bro63}
W.F. Brown, Thermal fluctuations of a single-domain particle, Phys.Rev. 130 (1963) 1677-1686.
\bibitem{Bro79}
W.F. Brown, Thermal fluctuations of fine ferromagnetic particles, IEEE Transactions on Magnetics, 15 (1979) 1196-1208.
\bibitem{Cof12}
W.T. Coffey, Y.P. Kalmykov, Thermal fluctuations of magnetic nanoparticles: Fifty years after Brown,
J. Appl. Phys. 112 (2012) 121301.
\bibitem{Cof94}
W.T. Coffey, D.S. F. Crothers, Yu.P. Kalmykov, E.S. Massawe, J.T. Waldron, Exact analytic formula for the correlation time of a single-domain ferromagnetic particle, Phys.Rev. E 49 (1994) 1869-1882.
\bibitem{Cof98}
W.T. Coffey, Finite integral representation of characteristic times of orientational
relaxation processes: Application to the uniform bias force effect in relaxation
in bistable potentials. Adv. Chem. Phys. 103 (1998) 259-333.
\bibitem{Cof961}
W.T. Coffey, D.S.F. Crothers, Comparison of methods for the calculation of superparamagnetic relaxation times, Phys. Rev. E 54 (1996) 4768-4774.
\bibitem{Cof001}
W.T. Coffey, D.S.F. Crothers, S.V. Titov, Escape times for rigid Brownian rotators in a bistable potential from the time evolution of the Green function and the characteristic time of the probability evolution, Physica A 298 (2001) 330-350.
\bibitem{Xia01}
J. Xiang, J. Sun, N.S. Sampson, The importance of hinge sequence for loop function
and catalytic activity in the reaction catalyzed by triosephosphate isomerase, J. Mol. Biol. 307 (2001) 1103-1112.
\bibitem{Kok12}
M. Kokkinidis, N.M. Glykos, V.E. Fadouloglou, Protein  flexibility and enzymatic catalysis,
Advances in protein chemistry and structural biology, 87 (2012) 181-218.
\bibitem{Pas92}
R.W. Pastor, A. Szabo, Langevin dynamics of a linear rotor in a Maier–Saupe potential: Kramers turnover of the flipping rate, J. Chem. Phys. 97 (1992) 5098-5112.
\bibitem{Ron95}
Heun Equations, A. Ronveaux (ed.), Oxford Univ. Press, 1995.
\bibitem{Sla00}
S.Y. Slavyanov, W. Lay, Special functions, a unified theory based on singularities, Oxford: Oxford Mathematical Monographs, 2000.
\bibitem{Fiz12}
P.P. Fiziev, D.R. Staicova, Solving systems of transendental equations involving the Heun functions,
AJCM 2 (2012) 95-105.
\bibitem{Fiz10}
P.P. Fiziev, Novel relations and new properties of confluent Heun functions and
their derivatives of arbitrary order, J. Phys. A: Math. Theor. 43 (2010) 035203.
\end{thebibliography}
\end{document}